# Locally addressable tunnel barriers within a carbon nanotube

M. J. Biercuk, N. Mason, J. M. Chow and C. M. Marcus

*Department of Physics, Harvard University,*
*Cambridge, Massachusetts 02138*

We report the realization and characterization of independently controllable tunnel barriers within a carbon nanotube. The nanotubes are mechanically bent or kinked using an atomic force microscope, and top gates are subsequently placed near each kink. Transport measurements indicate that the kinks form gate-controlled tunnel barriers, and that gates placed away from the kinks have little or no effect on conductance. The overall conductance of the nanotube can be controlled by tuning the transmissions of either the kinks or the metal-nanotube contacts.

Carbon nanotubes are a leading material system for molecular electronic device applications as well as for fundamental studies of the electronic properties of low-dimensional systems. Single-walled carbon nanotubes can function as nanoscale analogues of electronic elements such as field-effect transistors[1-5] and interconnects[6, 7] in integrated circuits. In addition, nanotubes behave as ballistic conductors with large current density capacity, and also display single electron (quantum dot) charging and quantum effects at low temperatures[8-13]. Conformational changes in nanotubes modify their electronic properties: devices such as diodes have been created using intra-tube tunnel barriers formed from mechanically induced defects or "kinks"[14-16]. Scanned-gate microscopy indicates that the kinks serve as scattering centers[8, 16], consistent with theoretical predictions for bending defects in nanotubes[17, 18]. The properties of separate tunnel barriers were not independently adjustable in previous studies, however, as only a single backgate was available.

In this Letter we describe the realization of locally addressable tunnel barriers within a single carbon nanotube. The barriers are due to bending defects, formed by mechanically kinking[14] the nanotube. We show that electrostatic gates placed near each of the kinks independently tune these tunnel barriers from transparent to opaque, whereas gates placed away from the kinks (over undeformed sections of the nanotube) have little or no effect. In addition, a global backgate is used to modify the transparency of the metal-nanotube contacts.

Carbon nanotubes were grown via chemical vapor deposition from patterned Fe catalyst islands on a degenerately doped $Si/SiO_2$ wafer, using methane as the carbon source[19]. After locating a target nanotube with an atomic force microscope (in conventional raster mode), the cantilever tip was lowered and then used to push the nanotube laterally under computer control. The resulting kinks had slightly different shapes, with typical lateral deviations of ~200 nm from their undisturbed (i.e., straight) configuration. After kinking, the nanotubes were contacted with Ti/Au electrodes, forming devices of length ~ 1-2 μm. The wafers were then coated (without patterning) with ~125 nm SiO$_2$ deposited via plasma-enhanced chemical vapor deposition at room temperature[20]. Multiple Cr/Au top gates (~ 200-300 nm across, ~ 40 nm thick), patterned using electron beam lithography and lift-off, were positioned over each tube, with a gate placed near each kink and at least one additional gate over an unkinked section of the same nanotube as a control. (Additionally, the doped Si substrate was used as a global backgate).

We first discuss a device with one kink and three top gates, one gate near a kink (G2) and two gates over unkinked sections of the nanotube (G1 and G3), as shown in the inset to Fig. 1a. Two-terminal conduction is measured from 1.7K to 300K using a voltage bias of 10 mV and measuring dc current using an Ithaco 1211 current amplifier and digital voltmeter. All gates are connected to dc voltage sources, and are set to 0V (relative to tube ground) unless otherwise specified. Measurements of conductance as a function of backgate voltage, $V_{BG}$, at room temperature and ~135 K show field-effect behavior consistent with p-type doping. Setting $V_{BG} = 0V$ and applying a voltage, $V_{G2}$, to G2 gives a strong gate response, while applying voltages $V_{G1}$ and $V_{G3}$ to gates G1 and G3 produces little change in conductance (Fig. 1a). These results, as well as those below, demonstrate the local effect of the top gates. The lack of gate response to G1 and G3 demonstrates the field-effect behavior observed as a function of $V_{BG}$ is not due to a depletion of the bulk of the nanotube. Rather, the dominant effect of the backgate is presumably to alter tunneling at the nanotube-metal contacts[21].

At low temperature, regions of charges confined by one or more kinks give rise to Coulomb blockade oscillations in conductance as a function of various gate voltages. A sweep of $V_{G2}$ at 1.7 K reveals widely spaced Coulomb blockade peaks (Fig. 1b) even at large bias voltage (10!mV). Similar single-electron charging events are observed up to ~135 K. Device conductance below ~30 K is zero whenever $V_{G2} = 0V$ regardless of other gate values, suggesting that the single kink acts as a small quantum dot (i.e., a device with at least *two* barriers) with a large charging energy (~20meV). Previous experiments have established that closely spaced mechanical defects can create such quantum dots[15, 16]. With $V_{G2}$ tuned to a conductance peak, Coulomb oscillations are seen as a function of either $V_{G3}$ or $V_{G1}$ on a gate-voltage scale considerably smaller than that measured for $V_{G2}$, likely due to a smaller charging energy (Inset Fig. 1b, Fig. 1c). This is consistent with larger quantum dots forming between the defect and tunnel barriers at the contacts[22, 23], with energy levels modulated by $V_{G1}$ and $V_{G3}$. In this case, resonant transport occurs only when the energy levels between all dots (source-kink, kink, kink-drain) are resonantly aligned. At $V_{G1}=0V$ the source-kink dot is transparent; the resulting double-dot system is evident in a two-dimensional plot of dc current as a function of $V_{G2}$ and $V_{G3}$ (inset to Figure 1b). The pattern of alternating high and low conductance regions is due to energy levels moving in and out of resonant alignment between the two dots[24].

We next discuss a device with multiple kinks, each with a nearby gate. The inset to Figure 2 shows a realistic schematic of a device with two gated kinks (under G4 and G6) and an additional top gate (G5) over the section of the nanotube between the kinks.

Room temperature measurements show a strong response to $V_{BG}$ (Fig. 2a). When $V_{BG}$ is swept at 70!K, the field-effect behavior is superposed with irregular Coulomb oscillations. A two-dimensional plot of dc current ($V_{SD}$=10mV) as a function of $V_{BG}$ and $V_{G5}$ shows the negligible effect of $V_{G5}$ (Fig. 2b). In contrast, measurements of dc current as a function of $V_{BG}$ and $V_{G4}$ show a strong gate response to $V_{G4}$. The nearly horizontal features visible in Fig. 2c correspond to values of $V_{G4}$ at which transport is suppressed by reducing the transparency of the underlying kink. The dashed line points out one such feature, which has a slight slope because of capacitive coupling between the backgate and the kink. Oscillatory features as a function of $V_{BG}$ persist, again with a small slope (dotted line) due to coupling between $V_{G4}$ and $V_{BG}$.

Local control over individual kinks is demonstrated in Figure 3a, where dc current ($V_{SD}$=10mV) is plotted as a function of $V_{G4}$ and $V_{G6}$. To maximize conductance we set $V_{BG}$=-10V. When both $V_{G4}$ and $V_{G6}$ are large and negative, current through the device is large, indicating that both kinks are transparent. Beyond this corner of the plot, conductance is strongly suppressed by either of the two top-gate voltages. The sharp turn-on of the double-kink device in response to multiple input voltages is equivalent to an AND logic element. The appearance of perpendicular horizontal and vertical bands indicates that G4 and G6 have very little cross-capacitance. Measurements at smaller negative backgate voltages produce similar features with a suppressed current flow. Measurements of dc current as a function of $V_{G4}$ and $V_{G5}$ produce the same features in response to $V_{G4}$, but show no response to $V_{G5}$ (Fig. 3b). Despite differences in the shapes of the three kinks studied here, we note that all respond to gate voltages on approximately the same scale.

In conclusion, we have fabricated and investigated nanotube devices with intentional bend and kinks created using an atomic force microscope, and electrostatic top gates near the kinks. These kinks behave as controllable tunnel barriers with local gate addressability. In contrast, gates placed away from the kinks on the same nanotubes had little or no effect on conductance. Future devices aim to use gate-tunable kinks to manipulate electron hybridization between different sections of a nanotube in the quantum regime.

This work was supported by funding from the NSF under EIA-0210736, the Army Research Office, under DAAD19-02-1-0039 and DAAD19-02-1-0191, and the Harvard MRSEC. M.J.B. acknowledges support from an NSF Graduate Research Fellowship and from an ARO Quantum Computing Graduate Research Fellowship. N.M. acknowledges support from the Harvard Society of Fellows.

Figure Captions:

Fig. 1 (a) Gate response of single-kink device. Plot shows dc current as a function of three gate voltages, $V_{G1}$, $V_{G2}$ and $V_{G3}$ over different sections of the nanotube, measured at T=135K (bottom, left axes) and of $V_{BG}$ at room temperature (top, right axes). Inset: SEM image of the device with small arrows indicating the position of the nanotube under the $SiO_2$ insulating layer and large arrowhead indicating the location of the kink.

Fig. 1(b) Gate response of kinked device to $V_{G2}$ (over kink) at low temperature showing Coulomb charging. Inset: Grayscale plot of dc current as a function of $V_{G2}$ and $V_{G3}$ displaying resonant transport through multiple dots in series.

Fig. 1(c) Horizontal slices through grayscale plot in Fig. 1(b) at on- and off-peak values for $V_{G2}$. On peak shows coulomb blockade as a function of $V_{G3}$ while off-peak shows total suppression of current.

Fig. 2(a) Room temperature conductance of double-kink device displaying approximate p-type field-effect behavior. Inset: Schematic of double kink device, showing kinks of different shapes and bending angles.

Fig. 2(b) Grayscale plot of dc current as a function of $V_{BG}$ and $V_{G5}$. Coulomb oscillations appear in $V_{BG}$, but no effect of $V_{G5}$ is evident.

Fig. 2(c) dc current as a function of $V_{BG}$ and $V_{G4}$ showing the ability to tune transport through gate control of a kink. Two scales of voltage additivity are present, indicated by dotted and dashed lines (see text).

Fig. 3(a) dc current as a function of $V_{G4}$ and $V_{G6}$, both over kinks in the nanotube, demonstrating the ability to inhibit transport through the device by appropriately tuning either kink.

Fig. 3(b) dc current as a function of $V_{G5}$ and $V_{G4}$. $V_{G5}$ has essentially no effect on transport through the device.

Figure 1

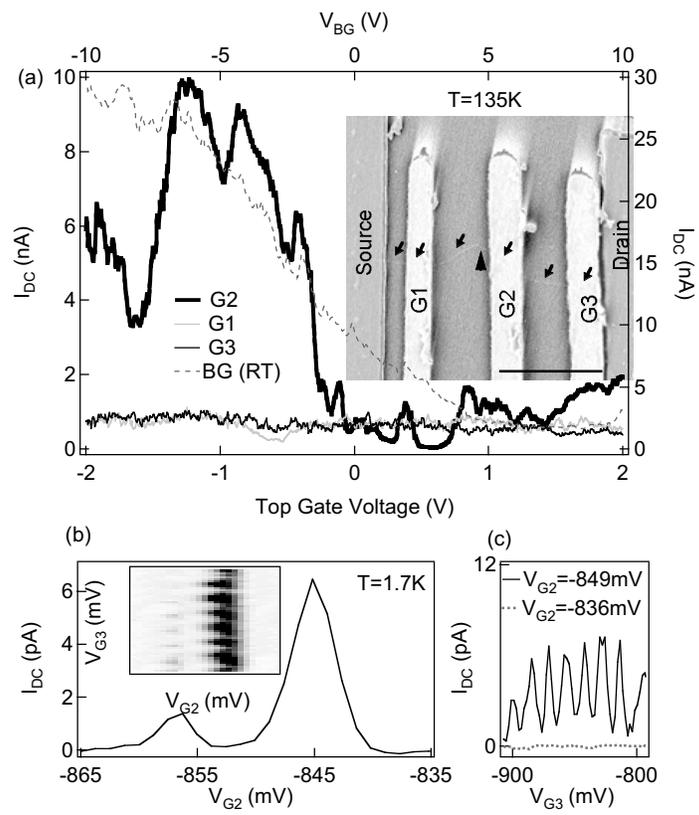

Figure 2

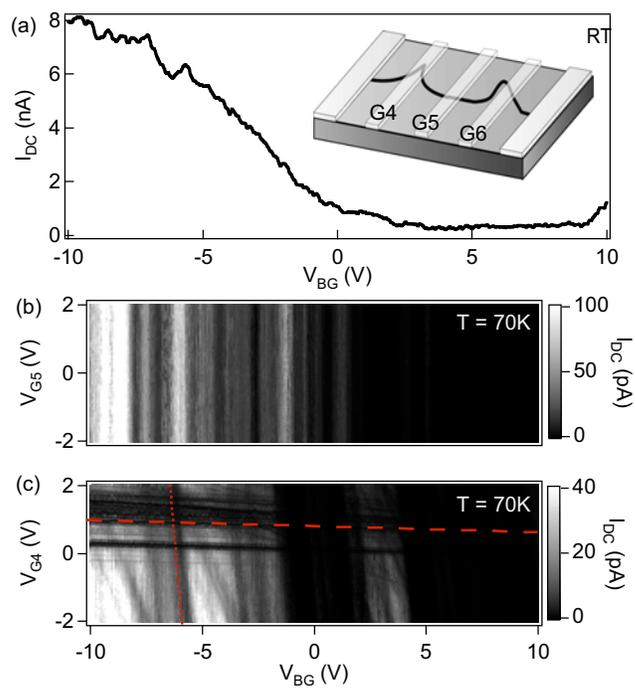

Figure 3

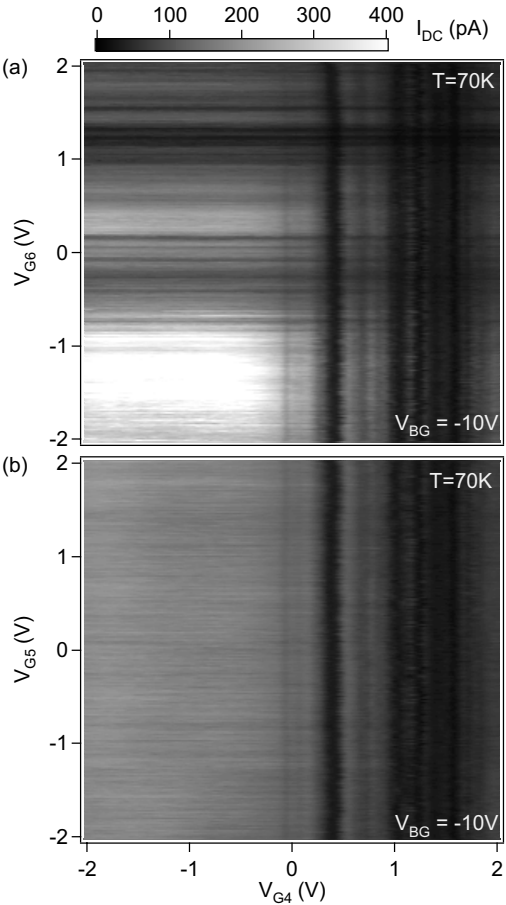